\documentclass{elsart}

\usepackage[english]{babel}
\usepackage{dcolumn}
\usepackage{amsmath,amssymb,epsfig,float}

\begin{document}
\begin{frontmatter}
\title{First law of thermodynamics and Friedmann-like equations in braneworld cosmology}
\author[apctp]{Xian-Hui Ge}~~~
\address[apctp]{Asia-Pacific Center for Theoretical Physics,\\
 Pohang 790-784, Republic of Korea }
\begin{abstract}
We derive the Friedmann-like equations in braneworld cosmology by
imposing the first law of thermodynamics and Bekenstein's
area-entropy formula on the apparent horizon of a
Friedmann-Robertson-Walker universe in both Randall-Sundrum II
gravity and Dvali-Gabadadze-Porrati gravity models. Israel's
boundary condition plays an important role in our calculations in
both cases, besides the first law of thermodynamics and Bekenstein's
area-entropy formula. The results indicate that thermodynamics on
the brane world knows the behaviors of gravity.\\


\end{abstract}
\end{frontmatter}

\maketitle

\section{Introduction}
\hspace*{7.5mm}The relations between thermodynamics of space time
and the nature of gravity is one of the most intriguing topics in
theoretical physics. The four laws of black hole thermodynamics were
first derived from the classical Einstein equation \cite{bar}. With
the discovery of Hawking radiation of black holes \cite{haw}, it
became clear that the analogy is actually an identity\cite{bek}. On
the other hand, the Einstein equation is derived by Jacobson from
the proportionality of entropy and horizon area together with the
first law of thermodynamics $\delta Q=TdS$\cite{jac}. Verlinde found
that
 the Friedmann equation in a radiation dominated Friedmann-Robertson-Walker (FRW) universe
 can be written in an analogous form of the Cardy-Verlinde formula,
 an entropy formula for a conformal field theory\cite{ver}. The
 above observations imply that thermodynamics of space time and the
 Einstein equation are closely related \cite{pad}.
  In ref.\cite{cai}, Cai and Kim derived the Friedmann equations of
FRW universe with any spatial curvature by applying the first law of
thermodynamics to the apparent horizon and assuming the geometric
entropy is given by a quarter of the apparent horizon area.\\
\hspace*{7.5mm}The main interest of this paper relys on the
relations between the first law of thermodynamics and the
Friedmann-like equation in braneworld cosmology. In the study of
three-brane cosmological models, an unusual law of cosmological
expansion on the brane has been reported (for incomplete references
see \cite{bin,bin2,col,barc,sht}). According to this law, the energy
density of matter on the brane quadratically enters the right-hand
of the new Friedmann equation for the braneworld cosmology, in
contrast with the standard cosmology, where the Friedmann equation
depends linearly on the energy density of matter. In this paper, we
shall derive the Friedmann-like equation in the braneworld cosmology
from the first
law of thermodynamics.\\
\hspace*{7.5mm}This paper is organized as follows. In section 2, we
discuss the relations between thermodynamics and gravity in
Randall-Sundrum II (RSII) model \cite{randall}. In section 3, we
derive the Friedmann-like equation in Dvali-Gabadadze-Porrati (DGP)
model \cite{dav} by applying the first law of thermodynamics to the
apparent horizon together with Israel's boundary condition. We
present our conclusions in section 4.
\section{FRW universe in Randall-Sundrum II gravity}
\hspace*{7.5mm}We consider a D-brane located on the boundary of a
$(n+1)$-dimensional Anti-de Sitter space time, as our visible
universe with ordinary matter being trapped on this brane by string
theory effects. We will derive the braneworld cosmological evolution
equations from the first law of thermodynamics. At each point of on
the brane, we define a space-time unit normal, $N_A=N_A(x)$, to the
surface that satisfies $g^{AB}N_AN_B=1$. $g^{AB}$ is the bulk metric
and the indices $A,B$ run over all the bulk coordinates. The bulk
metric induces a metric on the brane,
\begin{equation}
g_{\mu\nu}=g_{AB}-N_{AB}.
\end{equation}
The Einstein equation on the 3-brane world has been derived
in\cite{meada}.  The Einstein equation on this $(n-1)$-brane is
given by\cite{cao,padilla}(we use the notations in \cite{cao}),
\begin{equation}
\label{einstein} ^{(n)}G_{\mu\nu}=-\Lambda_{n}g_{\mu\nu}+8\pi
G_{n}T_{\mu\nu}+\kappa_{n+1}^4\Pi_{\mu\nu}-E_{\mu\nu},
\end{equation}
where
\begin{eqnarray}
\label{Lam}
\Lambda_{n}&=&\kappa_{n+1}^2\left[\frac{n-2}{n}\Lambda_{n+1}+\frac{(n-2)}{8(n-1)}\kappa_{n+1}^2\lambda^2\right],\\
\label{G} G_{n}&=&\frac{n-2}{32\pi(n-1)}\lambda\kappa_{n+1}^4,\\
\label{pi}
\Pi_{\mu\nu}&=&-\frac{1}{4}T_{\mu\alpha}T_{\nu}^{\alpha}+\frac{1}{4(n-1)}TT_{\mu\nu}+\frac{1}{8}g_{\mu\nu}T_{\alpha\beta}T^{\alpha\beta}
-\frac{1}{8(n-1)}T^2g_{\mu\nu},
\end{eqnarray}
where $\kappa_{n+1}^2=8\pi G_{n+1}$, $\Lambda_{n}$ is the brane
cosmological constant, $\lambda$ is the brane tension, $G_n$ is the
Newton's constant in n-dimensions, $T_{\mu\nu}$ is the
energy-momentum tensor of matter in the brane world, and
$E_{\mu\nu}$ is the Weyl tensor, which is vanishing in pure Anti-de
Sitter space time. We can see from (\ref{einstein}) that the
n-dimensional effective equations of motion, being given entirely in
terms of quantities defined on the brane, is independent of the
evolution of the bulk space time. Different from the standard
Einstein's equation in general relativity, the energy-momentum
tensor in the right-hand of (\ref{einstein}) is modified due to the
presence of extrinsic curvature when project $(n+1)$-dimensional
bulk quantities to the brane. We will not solve (\ref{einstein}) in
the rest of our work, but just utilize the modified energy-momentum
tensor in the right-hand of (\ref{einstein}) together with the first
law of thermodynamics to derive cosmological evolution equation on
the brane.\\
\hspace*{7.5mm}In the following calculation, we assume the
cosmological constant $\Lambda_n$ vanishes, since we are mainly
concern with the dynamical aspects of the braneworld cosmology while
$\Lambda_n$ is a constant and does not vary with time. The induced
metric on the brane is the FRW metric with the form,
\begin{equation}
\label{bmetric}
ds^2_{n}=-d\tau^2+a^2(\tau)\left(\frac{dr^2}{1-kr^2}+r^2d\Omega^2_{n-2}\right)
\end{equation}
where $d\Omega^2_{n-2}$ is the metric on an $(n-2)$-dimensional
Euclidean unit space of constant curvature, $k=1,0,-1$ which
corresponds to the unit sphere, plane, and hyperboloid respectively.
The metric (\ref{bmetric}) can be rewritten as\cite{bak}
\begin{equation}
\label{metric}
 ds^2=h_{ab}dx^adx^b+\tilde{r}^2d\Omega^2_{n-2},
\end{equation}
where $\tilde{r}=a(\tau)r$ and $x^0=\tau$, $x^1=r$. The
2-dimensional metric
$h_{ab}={\rm{diag}}(-1,\frac{a^2(\tau)}{1-kr^2})$. Since the
apparent horizon satisfies the equation
$h^{ab}\partial_{a}\tilde{r}\partial_{b}\tilde{r}=0$, we obtain the
radius of the apparent horizon,
\begin{equation}
\tilde{r}_{A}=\frac{1}{\sqrt{H^2+\frac{k}{a^2}}},
\end{equation} where $H=\frac{\dot{a}}{a}$ is the Hubble parameter.
For a dynamical space time, the apparent horizon is regarded as the
horizon satisfying the Bekenstein  area-entropy formula\cite{hay}. A
unified law of black hole dynamics and relativistic thermodynamics
is derived in spherically symmetric general relativity, where the
gradient of the active gravitational energy $E$ determined by the
Einstein's equation is divided into\textit{ energy-supply} and
\textit{work} terms\cite{hay}. Assume the matter on the brane is
given by a homogeneous perfect fluid of density $\rho$ and pressure
$p$, so that
\begin{equation}
T_{\mu\nu}=(\rho+p)U_{\mu}U_{\nu}+pg_{\mu\nu}.
\end{equation}Substituting the tensor back to Eq.(\ref{einstein})
and keep in mind that $\Lambda_{n}$ and $E_{\mu\nu}$ are vanishing,
we find that,
\begin{eqnarray}
\label{tensor}\tilde{T}^{~\mu}_{\nu}&=&T^{~\mu}_{\nu}+\frac{\kappa_{n+1}^4}{\kappa_{n}^2}\Pi^{~\mu}_{\nu}\nonumber\\
&=&{\rm diag}\left(-\rho-
\frac{n-2}{8(n-1)}\frac{\kappa_{n+1}^4}{\kappa_{n}^2}\rho^2,p+\frac{n-2}{8(n-1)}
\frac{\kappa_{n+1}^4}{\kappa_{n}^2}\rho(\rho+2p),...\right),
\end{eqnarray}where $\tilde{T}^{\mu}_{\nu}$ can be regarded as  the total effective
energy-momentum tensor on the brane and $\kappa_{n}^2=8\pi G_{n}$
denotes  Newton's constant.
 Now we would like to project the total effective
energy-momentum tensor $\tilde{T}^{\mu}_{\nu}$ to the normal
direction of the $(n-1)$-space and have it denoted as $T^{ab}$.
Then, one may define the \textit{work density} by \cite{bak,hay},
\begin{equation}
\label{work} W=-\frac{1}{2}T^{ab}h_{ab},
\end{equation}
and the \textit{energy-supply} vector is given by,
\begin{equation}
\label{energy-supply}
 \Psi_{a}\equiv
T^{~b}_{a}\partial_{b}\tilde{r}+W\partial_{a}\tilde{r}
\end{equation} As noted in ref.\cite{hay}, the work density at the
apparent horizon may be viewed as the work done by the change of the
apparent horizon and the energy-supply at the horizon is total
energy flow through the apparent horizon. Thus the total change of
energy on the apparent horizon can be written as,
\begin{equation}
\label{unified} \nabla E=A\Psi+W\nabla V,
\end{equation}where $A=(n-1)\Omega_{n-1}\tilde{r}^{n-2}$ and $V=\Omega_{n-1}\tilde{r}^{n-1}$ are the area and the volume
of the $(n-1)$-brane with radius $\tilde{r}$, and
$\Omega_{n-1}=\pi^{(n-1)/2}/\Gamma((n-1)/2+1)$ is the volume of an
$(n-1)$-dimensional unit ball. Eq.(\ref{unified}) is interpreted as
\textit{unified first law}\cite{hay}.\\
\hspace*{7.5mm}Let us turn to calculating the heat flow $\delta Q$
through the apparent horizon during an infinitesimal time interval
$dt$, while keep the volume of the brane stable, namely $\nabla
V=0$. In this sense,\begin{equation} dE\equiv A\Psi.\end{equation}
The heat flow $\delta Q$ through the apparent horizon is just the
amount of energy crossing it during the time interval $dt$. Thus,
$\delta Q=-dE$ is the change of the energy inside the apparent
horizon. From (\ref{tensor}), (\ref{work}) and
(\ref{energy-supply}), we obtain the expression of work density and
energy-supply,
\begin{eqnarray}
W&=&-\frac{1}{2}(-\rho+p+\frac{n-2}{4(n-1)}\frac{\kappa_{n+1}^4}{\kappa_{n}^2}\rho
p),\\
\Psi_{t}&=&-\frac{1}{2}(\rho+p)\dot{a}r-\frac{n-2}{8(n-1)}\frac{\kappa_{n+1}^4}{\kappa_{n}^2}\rho(\rho+p)\dot{a}r,\\
\Psi_{r}&=&\frac{1}{2}(\rho+p)a+\frac{n-2}{8(n-1)}\frac{\kappa_{n+1}^4}{\kappa_{n}^2}\rho(\rho+p)a
\end{eqnarray}
We assume the entropy of the apparent horizon is described by the
Bekenstein area formula,
\begin{equation}
S=\frac{A}{4G_{n}},
\end{equation}
with the horizon temperature,
\begin{equation}~T=\frac{1}{2\pi\tilde{ r}_{A}}. \end{equation}
 Calculating the amount of energy crossing the
apparent horizon and applying the first law, $-dE=A\Psi dt= TdS$, we
find
\begin{equation}
\label{friedmann}
\kappa_{n}^2(\rho+p)\left(1+\frac{n-2}{8(n-1)}\frac{\kappa_{n+1}^4}{\kappa_{n}^2}\rho\right)=-(n-2)(\dot{H}-\frac{k}{a^2}).\end{equation}
Integrating (\ref{friedmann}) together with the energy conservation
equation on the brane,
\begin{equation}
\dot{\rho}+(n-1)H(\rho+p)=0,
\end{equation}
we finally obtain the cosmological evolution equation in the
Randall-Sundrum II universe
\begin{equation}
\label{friedmanone}H^2+\frac{k}{a^2}=\frac{2\kappa_{n}^2}{(n-1)(n-2)}\rho+\frac{\kappa_{n+1}^4}{4(n-1)^2}\rho^2
.\end{equation} The above equation is exactly the Friedmann-like
equation that describes the cosmological evolution of
Randall-Sundrum II universe, where the cosmological constant is set
to be vanishing here.\\
\hspace*{7.5mm}One should note that in the above calculations, we
have simply use the Bekenstein entropy-area formula. In
ref.\cite{cao,cai2}, the authors started from the Friedmann-like
equation on the braneworld together with the first law of
thermodynamics $-dE=TdS+WdV$ and then derived the entropy
expressions for the apparent horizon. The results show that the
apparent horizon is proportional to the horizon area but with some
corrections and only in the large horizon radius limit, the entropy
obeys the $1/4$ area law. Here, we would like to justify the
assumption about the usage of Bekenstein formula. In the above
calculation, we have assumed during the time interval $dt$, the
volume of the brane does not change, namely $dV\propto
d\tilde{r}_{A}=0$. In other words, we have only considered the
isometric process, $-dE=TdS$. Thus, if one starts from the
Friedmann-like equation (\ref{friedmanone}) and consider the
isometric thermodynamical  process, one can find that the apparent
horizon entropy obeys the $1/4$ area law. In this sense, let us
examine the expression of the surface gravity at the apparent
horizon, which is generally given by,
\begin{eqnarray}
\label{k} \kappa
=\frac{1}{2\sqrt{-h}}\partial_{a}(\sqrt{-h}h^{ab}\partial_{b}\tilde{r})
=-\frac{1}{\tilde{r}_{A}}\left(1-\frac{\dot{\tilde{r}}_A}{2H\tilde{r}_{A}}\right).
\end{eqnarray}
The second term in the parenthesis is vanishing while
$d\tilde{r}_{A}=\frac{d\tilde{r}_{A}}{dt}dt=0$. As a consequence,
the temperature is read as $T=\frac{1}{2\pi \tilde{r}_{A}}$.
Therefore, an isometric thermodynamical process on the apparent
horizon encodes the cosmological evolution equation. However, as
suggested in ref.\cite{cao,cai2}, this does not mean that the
Bekenstein area formula used in the calculations above really reveal
the relation between horizon entropy and horizon geometry, since the
presence of an extra dimension cause the gravity here to be
non-Einstein type. More explicit entropy expressions of the apparent
horizon of
FRW universe in RSII model are given in ref.\cite{cao}.\\
\hspace*{7.5mm}In summary, If the entropy and temperature of the
brane can be described by the apparent horizon, an isometric
thermodynamical process on the apparent horizon should ensure the
Bekenstein area formula established and  from the first law of
thermodynamics $-dE=TdS$, we are able to obtain the Friedmann-like
equation for an $(n+1)$-dimensional Randall-Sundrum II universe with
arbitrary spatial curvature.
\section{ FRW universe in Dvali-Gabadadze-Porrati gravity}
 \hspace*{7.5mm}In this section, we explore the relations between the first law of
 thermodynamics and cosmology
based on the Dvali-Gabadadze-Porrati gravity model\cite{dav}. In
this model, a 3-brane is embedded in a spacetime with an
infinite-size extra dimension. The usual gravitational laws is
obtained by adding to the action of the brane an Einstein-Hilbert
term computed with the intrinsic curvature on the brane.
Particularly, one recovers a standard four-dimensional (4D)
Newtonian potential for small distances, whereas gravity is in a 5D
regime for large distances. The cosmology of this model in the case
of a 5D bulk was studied by Bin$\rm \acute{e}$truy et
al\cite{bin,bin2,dav}. It is shown there that if the cosmological
model contains a scalar curvature term in the action for the brane,
besides the brane and bulk cosmological constraints, the presence of
the scalar curvature term in the brane action can lead to a
late-time acceleration of the universe even in the absence of any
material form of dark energy \cite{deff}.\\
\hspace*{7.5mm}The cosmology in a 5-dimensional bulk is described by
the metric,
\begin{equation}
\label{bulkmetric}
ds^2=-n^2(\tau,y)d\tau^2+a^2(\tau,y)\left(\frac{dr^2}{1-kr^2}+r^2d\Omega^2_{2}\right)+b^2(\tau,y)dy^2,
\end{equation}where $d\Omega^2$ is the metric of a 2-dimensional Euclidean unit space and $k=0,\pm 1$. Since we mainly focus our
attention on the hypersurface defined by $y=0$\cite{bin}, which the
world volume of brane is identify with that of our universe, it is
important to examine the behavior of metric in the bulk. The metric
(\ref{bulkmetric}) can be rewritten as\cite{bak}
\begin{equation}
\label{metric}
 ds^2=h_{ab}dx^adx^b+\tilde{r}^2d\Omega^2_{2},
\end{equation}
where $\tilde{r}=a(\tau,y)r$ and $x^0=\tau$, $x^1=r$. The
3-dimensional metric
$h_{ab}={\rm{diag}}(-n^2(\tau,y),\frac{a^2(\tau,y)}{1-kr^2},b^2(\tau,y))$.
The apparent horizon is defined by\cite{hay},
\begin{equation}
\label{apparent} h^{ab}\partial_{a}\tilde{r}\partial_{b}\tilde{r}=0,
\end{equation}
which denotes the radius of the apparent horizon,
\begin{equation}
\label{rA}
\tilde{r}_A=\frac{1}{\sqrt{{\frac{\dot{a}^2}{a^2n^2}}+\frac{k}{a^2}-\frac{{a'}^2}{a^2b^2}}}.
\end{equation}
Since $a(\tau,y)$, $b(\tau,y)$ and $n(\tau,y)$ are functions of the
fifth dimension coordinate $y$, the above apparent horizon is
different from that of standard cosmology. Hereafter we choose
$n^2(y=0)=1$, which implies that the time on the brane corresponds
to standard cosmic time, and assume the fifth dimension is static
$\dot{b}=0$ and set $b=1$. The $5$-dimensional Einstein equations in
the bulk is given by,
\begin{equation}
\label{bulk} \tilde{G}_{AB}\equiv
\tilde{R}_{AB}-\frac{1}{2}\tilde{R}\tilde{g}_{AB}=\kappa_{5}^2
\tilde{T}_{AB},
\end{equation}
where $\kappa_{5}^2=8\pi G_{5}$, $\tilde{R}_{AB}$ is the
five-dimensional Ricci tensor and $\tilde{T}_{AB}$ is the
energy-momentum tensor of matter. The energy-momentum tensor can be
further decomposed into two parts,
 \begin{equation}
\label{t} \tilde{T}^{~A}_{B}=-\Lambda_{5}+T^{~A}_{B}\mid_{\rm
brane},
\end{equation}where $\Lambda_{5}$ is the $5$-dimensional bulk
cosmological constant and the second term is the energy-momentum
tensor of the brane at $y=0$, which is given by,
\begin{equation}
T^{~A}_{B}\mid_{\rm brane}=-\lambda
\delta_{B}^{~\mu}\delta_{\nu}^{~A}
g^{~\mu}_{\nu}\delta(y)+\delta(y)\rm{diag}(-\rho,p,p,p,0),
\end{equation}
where $\lambda$ corresponds to brane tension (or cosmological
constant) and the last term is the matter content on the brane
($y=0$), which is assumed in the form of a perfect fluid for
homogenous and isotropic universe on the brane. We emphasize that
the brane tension can in principle be tuned to be zero
\cite{dav,gaba}. In view of the brane tension do not vary with
cosmic time $t$ and the dynamical aspects of the braneworld
cosmology are our object, hereafter  we follow the method of
ref.\cite{cai} and focus mainly  on the matter content of the brane
in the calculations. One will see lately that an integration
constant can be regarded as the brane tension, in case
 it
does not vanish.\\
  \hspace*{7.5mm}Now
we project the energy-momentum tensor of the matter content of the
brane to the normal direction of $(\tau,r,y)$ that is denoted as
$T^{~b}_{a}$. Note that we assume there are no flows of matter along
the fifth dimension and $T_{yy}\mid_{\rm brane}=0$. The work density
and energy-supply vector defined in (\ref{work}) and
(\ref{energy-supply}) here read as,
\begin{eqnarray}
W&=&\frac{1}{2}\delta(y)(\rho-p),\nonumber~~~
\Psi_{t}=-\frac{1}{2}\delta(y)(\rho+p)\dot{a}r,\nonumber\\
\Psi_{r}&=&\frac{1}{2}\delta(y)(\rho+p)a,~~~~
\Psi_{y}=\frac{1}{2}\delta(y)(\rho-p)a'r.
\end{eqnarray}
\hspace*{7.5mm}At the time interval $dt$, the amount of energy
crossing the apparent horizon is then,
\begin{equation}
\label{de}
 -dE=-A\Psi=\delta(y)A(\rho+p)H\tilde{r}_{A}dt.
\end{equation}
Since our discussions about the brane are confined to the
hypersurface $y=0$, and the volume of the brane is exactly that of
our observing universe, then considering an isometric
thermodynamical process across the apparent horizon, we can assume
that
 the entropy and temperature denoted by the apparent horizon can be
written as,
\begin{equation}
S=\frac{A}{4 G_{(4)} },~~~~ T=\frac{1}{2\pi \tilde{r}_{A}}
\end{equation}where $A=4\pi \tilde{r}_{A}^2$ is the  apparent horizon area.
Using the first law of thermodynamics, $-dE=TdS$, together with
(\ref{de}), we find that at $y=0$,
\begin{equation}
\label{fried}
\kappa_{4}^2(\rho+p)=-2(\dot{H}-\frac{k}{a^2}-\frac{a'\dot{a}'}{a\dot{a}}+\frac{{a'}^2}{a^2
}).
\end{equation}
The  energy conservation equation on the brane is given by,
\begin{equation}
\label{conservation} \dot{\rho}+3H(\rho+p)=0.
\end{equation}
From (\ref{fried}) and (\ref{conservation}), we finally obtain,
\begin{equation}
\label{friedman1} {H}^2+\frac{k}{a^2}-\frac{{a'}^2}{a^2
}=\frac{{\kappa_4}^2}{3}\rho+C,
\end{equation}where $C$ is the integration constant that denotes the brane
cosmological constant, namely $C=\Lambda_{4}$. The brane
cosmological constant is given by Eq.(3) with $n=4$,
\begin{equation}
\label{lamda}
\Lambda_{4}=\frac{\kappa_{5}^2}{6}(\Lambda_{5}+\frac{\kappa_{5}^2}{6}\lambda^2).
\end{equation}  The value of
$\frac{{a'}^2}{a^2 }$ can be determined by using the Israel junction
condition \cite{israel,meada,cao},
\begin{equation}
K^{+}_{\mu\nu}=K^{-}_{\mu\nu}=-\frac{{\kappa_{5}}^2}{2}\left(T_{\mu\nu}-\frac{1}{3}g_{\mu\nu}T\right),
\end{equation}where $K_{\mu\nu}=g_{\mu}^{\alpha}g_{\nu}^{\beta}\nabla_{\alpha}n_{\beta}$
denotes the extrinsic curvature of the brane, $T_{\mu\nu}$ is the
matter energy-momentum tensor of the brane, and the superscript
``+'' and ``-''refer to the ``+''side ($\rm lim_{y\rightarrow +0}$)
and ``-'' side ($\rm lim_{y\rightarrow -0}$) of the brane,
respectively. As a consequence, we have
\begin{equation}
\frac{{a'}^2}{a^2 }=\frac{{\kappa_{5}}^4}{36}\rho^2,
\end{equation} where the $Z_{2}$ symmetry $y\leftrightarrow -y$ has
been used here. Therefore, the cosmological evolution equation on
the brane has the final form,
\begin{equation}
\label{friedman2}
{H}^2+\frac{k}{a^2}=\frac{{\kappa_4}^2}{3}\rho+\frac{{\kappa_{5}}^4}{36}\rho^2+\Lambda_{4}.
\end{equation}
Eq.(\ref{friedman2}) is in agreement with Eq.(\ref{friedmanone} when
$n=4$ and a non-vanishing cosmological constant . By using
Eq.(\ref{lamda}) and the relation (\ref{G}) (see also
\cite{csaki,cline}),
\begin{equation}
\label{tension} \kappa_4^2=\frac{\kappa^4_5}{6}\lambda
\end{equation}
Eq.(\ref{friedman2}) can be written as
\begin{equation}
\label{binu}
{H}^2+\frac{k}{a^2}=\frac{\kappa_{5}^2}{6}\Lambda_{5}+\frac{\kappa_{5}^4}{36}(\rho+\lambda)^2,
\end{equation}
which is the exact results of \cite{bin2} with vanishing Weyl
tensor. If an intrinsic curvature term is taken into accounted, then
Israel's junction condition is modified and the value of
$\frac{{a'}}{a }$ is replaced by \cite{deffayet},
\begin{equation}
\frac{{a'}}{a
}=-\frac{{\kappa_{5}}^2}{6}\rho^2+\frac{{\kappa_{5}}^2}{2\kappa_{4}^2}\left(H^2+\frac{k}{a^2}\right).
\end{equation}
Substituting the above formula back to (\ref{friedman1}) and using
(\ref{lamda}) and (\ref{tension}), we obtain cosmological evolution
equation in DGP gravity,
\begin{equation}
\label{rc}
H^2+\frac{k}{a^2}=\left(\sqrt{\frac{\kappa_4^2}{3}\rho_{b}+\frac{1}{4r_{c}^2}}+\frac{1}{2r_{c}}\right)^2,
\end{equation} where $\rho_{b}=\rho+\lambda $, $r_{c}=\frac{{\kappa_{5}}^2}{2\kappa_{4}^2}$ and $\Lambda_{5}$ here has been set to be zero.
Eq.(\ref{rc}) is precisely the Friedmann-like equation for a FRW
universe in DGP gravity first found in ref.\cite{deffayet}, when
intrinsic curvature is added to the brane.\\
\hspace*{7.5mm}Therefore, to derive the Friedmann-like equations for
FRW universe in DGP gravity,  Israel's boundary condition is needed,
besides applying the first law of thermodynamics and Bekenstein's
entropy-area formula to the apparent horizon. The intrinsic
curvature term added to the brane action plays a crucial role in
obtaining (\ref{rc}).
\section{{Conclusions}}
\hspace*{7.5mm}In conclusion, we have presented an alternative
methods to derive cosmological evolution equations in braneworld
cosmology. Employing the first law of thermodynamics, $-dE=TdS$ and
Bekenstein's area-entropy formula to the apparent horizon of FRW
universe in RSII gravity and DGP gravity, we have obtained the
Friedmann-like equations for brane world cosmology. Comparing
section 2 with section 3, we find that Israel's boundary condition
plays an important role in determining the effective energy-momentum
tensor $\tilde{T}_{\mu\nu}$ or the apparent horizon $\tilde{r}_{A}$.
Israel's boundary condition is not directly used in section 2, since
we were  start with the brane world Einstein equation, but it is
indeed crucial to derive the effective energy-momentum tensor
$\tilde{T}_{\mu\nu}$ \cite{meada}. An isometric thermodynamical
process happened across the apparent horizon is assumed in the above
discussions, thus whether a more general thermodynamical process
yields the Friedmann-like equations or not worth further
investigations. The detailed thermodynamical properties of apparent
horizon in braneworld scenario are
investigated in ref.\cite{cao,cai2} and some new physics about the braneworld gravity and thermodynamics is found.\\
\hspace*{7.5mm}It has been proved that gravity knows thermodynamics,
when Bekenstein and Hawking found that entropy of black hole horizon
is proportional to its horizon area\cite{haw,bek}. Black hole
entropy and temperature can be regarded as purely geometric
quantities in that the horizon area and surface gravity is
determined by space-time geometry. The above analysis and references
mentioned\cite{jac,ver,pad,cai,hay}, also indicate that
thermodynamics knows gravity. Once the form of energy-momentum
tensor of the system is determined, one can derive the spacetime
geometric evolution equations from Eqs.(15) and (19). The results
obtained are certainly related to the holographic principle, since
the brane world cosmology models are closely related to the AdS/CFT
conjecture\cite{mal}. It would be worthwhile to study the
implications of above analysis to the holographic principle or
quantum gravity in the future.
 \ack
 The author would like to thank Prof.
S. P. Kim for his helpful discussions.


\begin{thebibliography}{10}
\bibitem{bar} J.~M. Bardeen, B. Carter and S.W. Hawking, 
             Comm. Math. Phys. {\bf31} (1973) 161
\bibitem{haw} S.~W. Hawking,
             Comm. Math. Phys. {\bf43} (1975) 199
\bibitem{bek} J.~D. Bekenstein,
           Phys. Rev.  {\bf D 7} (1973) 2333
\bibitem{jac} T. Jacobson,
      {Phys. Rev. Lett.} 75 (1995) 1260 [gr-qc/9504004]
\bibitem{ver} E. Verlinde,
          hep-th/0008140
\bibitem{pad} T. Padmanabhan,
            {Class. Quantum Grav. }, {\bf19} (2002) 5387 [gr-qc/0204019]; T. Padmanabhan,
                 {Phys. Rep.} {\bf 406} (2005) 49
                 [gr-qc/0311036]; A. Paranjape, S. Sarkar,  T. Padmanabhan ,
                 {Phys.Rev.,} {\bf D 74},104015 (2006) [hep-th/0607240];
                A. Mukhopadhyay,  T. Padmanabhan ,
                 {Phys.~Rev.}, {\bf D 74}  (2006) 124023  [hep-th/0608120];
            T. Padmanabhan and A. Paranjape,
            {Phys.~Rev. }  {\bf D75} (2007) 064004 [gr-qc/0701003] ;
            R. G.
              Cai and L. M. Cao,
                  {Phys. Rev. } {\bf D 75} (2007)
                 064008
\bibitem{cai} R. G. Cai and S. P. Kim,
             JHEP 02 (2005) 050
\bibitem{bin}~P.~Bin$\acute{e}$truy,~C.~Deffayet,and~D.~Langlois,
             { Nucl.~Phys.} {\bf B565} (2000) 269 [hep-th/9905012]
 \bibitem{bin2} ~P.~Bin$\acute{e}$truy,~C.~Deffayet,~U.~Ellwanger
                 and~D.~Langlois,
                 {~Phys. Lett. B} {\bf 477} (2000) 285
                [hep-th/9910219]
\bibitem{col}~H.~Collins and ~B.~Holdom,
                 {Phys.~Rev.} {\bf D62} (2000) 105009 [hep-ph/0003173]
\bibitem{barc}~C.~Barcel$\acute{o}$ and ~M.~Visser,
              {Phys.~Lett.} {\bf B 482} (2000) 183
              [hep-th/0004056]
\bibitem{sht}   ~Y.~V.~Shtanov,
                [hep-th/0005193]
\bibitem{randall}L.~Randall and R.~Sundrum,
                 {Phys. Rev. Lett.} 83 (1999) 4690
\bibitem{dav}    G.~Dvali, ~G.~Gabadadze,~M.~Porrati,
                {Phys. Lett.}
                ~{\bf B 485} (2000) 208   [hep-th/0005016]
\bibitem{meada} T. Shiromiza,~K. Maeda, and M.~Sassaki,
                 {Phys. Rev.} {\bf D 62} (2000) 349
                 [gr-qc/9910076]; A.~N. Aliev and A.~E.
                 Gumrukcuoglu,
                 {Class.~Quant.~Grav.} 21 (2004) 5081
\bibitem{cao} R. G. Cai and L. M. Cao,
             hep-th/0612144
\bibitem{padilla}~A.~Padilla,
             Ph.~D Thesis, hep-th/0210217.
\bibitem{bak}~D. Bak and S.~J. Rey,
             {Class. Quantum Grav.} {\bf 17} (2000) L83 [hep-th/9902173]
\bibitem{hay}~S.~A. Hayward, S. Mukohyama and M.~C. Ashworth,
             {Phys. Lett.} {\bf A 256} (1999) 347
                [gr-qc/9810006]; S.~A. Hayward,
                 {Class. Quantum Grav.}
               {\bf 15} (1998) 3147 [gr-qc/9710089]
\bibitem{cai2}  A.~Sheykhi, B.~Wang and R. G.
                 Cai,
                 hep-th/0701198;  A.~Sheykhi, B.~Wang and R. G.
                 Cai,
                  hep-th/0701261
\bibitem{deff}   ~C.~Deffayet,~ G.~Dvali and ~G.~Gabadadze,
              {Phys. Rev.}
                 {\bf D   65}, ( 2002) 044023;
                  ~C.~Deffayet,~S.~J.~Landau,~J.~Raux,~M.~Zaldarriaga
                  and~P.~Astier,
                   {Phys. Rev.} D {\bf 66} (2002) 024019  [astro-ph/0201164]
\bibitem{gaba} G.~Dvali and ~G.~Gabadadze, Phys.~Rev.~D {\bf 63}
               (2001) 065007 [hep-th/0008054]
\bibitem{israel} W.~Israel,
                   {Nuovo Cimento} {\bf B 44} (1966) 1
\bibitem{csaki}~C.~Cs$\rm \acute{a}$ki, M.~Graesser  and
               C.~Kolda,~J.~Terning,
               {Phys.~Lett.} {\bf B~462} (1999) 34  [hep-th/9906513]
\bibitem{cline}~J.~M.~Cline,~C.~Grosjean and ~G.~Servant,
             {Phys.~Rev.~Lett.} 83 (1999) 4245 [hep-ph/9906523]
\bibitem{deffayet}~C.~Deffayet,
                      {Phys. Lett.}  {\bf B 502} (2001) 199
\bibitem{mal} J.~M.~Maldacena,
              {Adv. Theor. Math. Phys.} \textbf{2} (1998)
                231 [hep-th/9711200]
\end{thebibliography}
\end{document}